\documentclass{vgtc}                          

\graphicspath{{figures/}{pictures/}{images/}{./}} 

\usepackage{times}                     

\usepackage{tabu}                      
\usepackage{booktabs}                  
\usepackage{lipsum}                    
\usepackage{mwe}                       
\usepackage{spverbatim}

\usepackage{mathptmx}                  
\usepackage[normalem]{ulem}

\newcommand{\cg}{\textsc{ChatGPT}\xspace}
\newcommand{\cgs}{\textsc{ChatGPT}s\xspace}
\newcommand{\cgthree}{\textsc{ChatGPT-3.5}\xspace}
\newcommand{\cgfour}{\textsc{ChatGPT-4}\xspace}
\newcommand{\h}{\textsc{Human}\xspace}
\onlineid{0}

\vgtccategory{Research}

\vgtcinsertpkg

\title{Understanding Why \textsc{ChatGPT} Outperforms \textsc{Human}s in Visualization Design Advice}

\author{Yongsu Ahn\thanks{e-mail: anyon@bc.edu}\\
\scriptsize Boston College%
\and Nam Wook Kim\thanks{e-mail:nam.wook.kim@bc.edu} \\
\scriptsize Boston College
}

\abstract{
    This paper investigates why recent generative AI models outperform humans in data visualization knowledge tasks. Through systematic comparative analysis of responses to visualization questions, we find that differences exist between two \textsc{ChatGPT} models and human outputs over rhetorical structure, knowledge breadth, and perceptual quality. Our findings reveal that \textsc{ChatGPT}-4, as a more advanced model, displays a hybrid of characteristics from both humans and \textsc{ChatGPT}-3.5. The two models were generally favored over human responses, while their strengths in coverage and breadth, and emphasis on technical and task-oriented visualization feedback collectively shaped higher overall quality. Based on our findings, we draw implications for advancing user experiences based on the potential of LLMs and human perception over their capabilities, with relevance to broader applications of AI. 
} 


\begin{document}



\maketitle

\section{Introduction}
\label{sec:intro}

Many data visualization practitioners are self-taught, acquiring design knowledge on the go and building their skills informally through online examples and other digital resources \cite{dvsurveys, esteves2022learned}. When faced with design decisions, they often rely on intuition shaped by prior experiences and observations \cite{choi2023vislab,parsons2021understanding}. Others seek feedback from peers or online communities, such as the Data Visualization Society, to gain fresh perspectives, validate their design choices, or challenge underlying assumptions \cite{choi2023vislab, luther2015structuring}.

Recent generative AI models trained on internet-scale datasets have shown strong capabilities in data visualization knowledge tasks \cite{bendeck2024empirical}---for example, identifying misleading designs \cite{alexander2024can, lo2024good} and helping novices interpret charts \cite{choe2024enhancing}. Chatbots powered by these models can serve as design and learning assistants, offering guidance to data visualization practitioners more efficiently than the traditional design process. A recent study \cite{kim2023good} demonstrated this potential by feeding real-world design questions and feedback requests from the VisGuides platform into ChatGPT. The results showed that the AI's responses were often comparable to, or even better than, those generated by humans.

While the previous study offers valuable initial insights into ChatGPT’s potential as a design assistant, it has several limitations. The evaluation was carried out by a small group of researchers, which may not reflect the perspectives of actual practitioners. Moreover, it did not examine the underlying reasons behind ChatGPT’s superior performance compared to human counterparts, leaving open questions about not just how well ChatGPT performs, but why and under what conditions it excels or falls short.

This paper takes a deeper look into the comparative analysis of ChatGPT and human responses to data visualization questions, with a focus on their specific characteristics that shape the overall quality of their answers. To guide this investigation, our study explores the following research questions:

\vspace{-0.5em}
\begin{itemize}
    \item \textbf{RQ1} (\textbf{Response Characteristics}): What semantic and syntactic differences exist between \textsc{Human} vs. \textsc{ChatGPT} responses to data visualization questions?
    \vspace{-0.5em}
    \item \textbf{RQ2} (\textbf{Response Quality}): On what quality dimensions, \textsc{ChatGPT} responses are perceived better than \textsc{Human} counterparts?
    \vspace{-0.5em}
    \item \textbf{RQ3} (\textbf{Response Quality $\sim$ Response Characteristics}): What aspects of response content and intrinsic quality are associated with perceived overall quality of responses?
    \vspace{-0.5em}
\end{itemize}
\vspace{-0.5em}
\section{Methods}
\label{sec:method}

\begin{figure}
    \vspace{-1.25em} \includegraphics[width=\columnwidth]{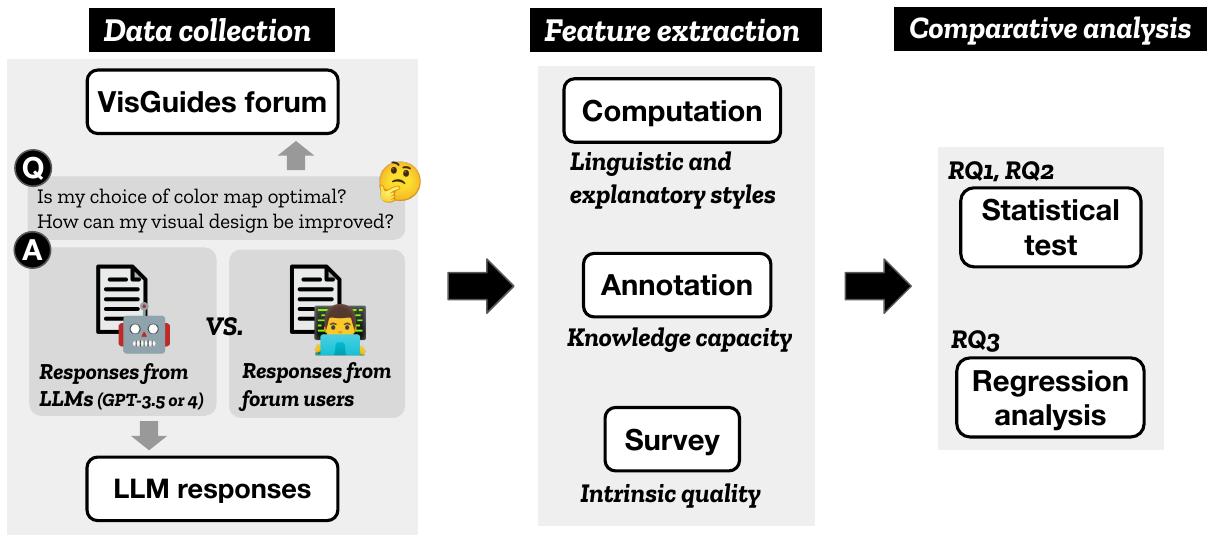}
    \vspace{-2.3em}
    \caption{\label{fig:method}
    The overview of the analysis pipeline. Our method facilitates the comparative analysis of three aspects of capabilities between human and generative AI.}
    \vspace{-1.75em}
\end{figure}

\begin{figure*}
    \vspace{-1.25em} \includegraphics[width=\textwidth]{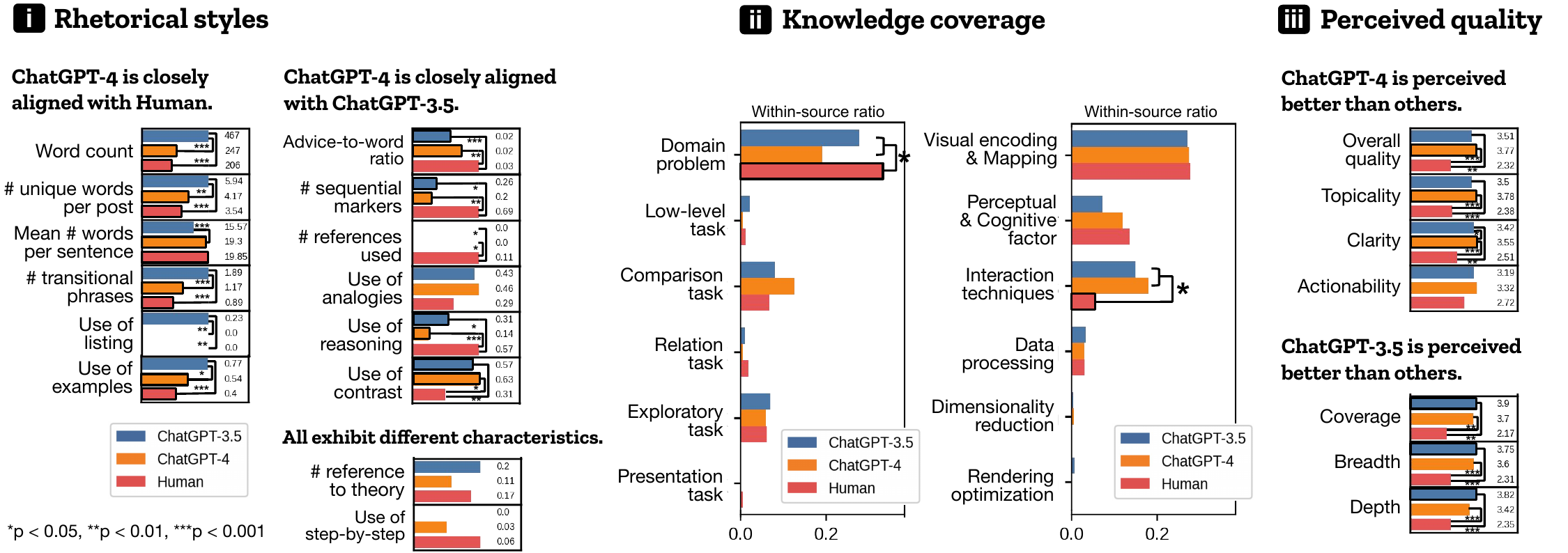}
    \vspace{-2.25em}
    \caption{\label{fig:features}
    Our analysis reveals distinct properties of responses from \h and \cgs over i) rhetorical styles, ii) knowledge coverage, and iii) perceived quality.}
    \vspace{-1.75em}
\end{figure*}

Figure \ref{fig:method} illustrates our methodological approach to address these questions through data collection, feature extraction for comparative analysis (RQ1), survey-based quality evaluation (RQ2), and regression modeling to identify quality-associated factors (RQ3).

\subsection{Data collection} For our analysis of visualization questions and responses, we used VisGuides \cite{diehl2018visguides}, a visualization-focused forum, where practitioners seek advice and feedback and respond to questions. We selected 119 of 226 questions based on inclusion criteria such as content sufficiency via a validation process from two coders’ unanimous agreement. To obtain responses from \cg, we fed each query to \cgthree in May 2023 and with images to \cgfour with Vision in May 2024. More details of the data collection can be found in \cite{kim2023good}.

\subsection{Feature selection and extraction} 
Drawing from existing research in social Q\&A and recent NLP studies, we selected 12 features to capture three categories of response characteristics, including rhetorical styles, knowledge coverage, and perceived quality.

The first category, \textbf{rhetorical styles}, examines the structural patterns exhibited in the responses. Studies in social Q\&A have investigated that a variety of quantitative features---including linguistic characteristics \cite{fu2015evaluating, li2015answer} or rhetorical behaviors \cite{shah2010evaluating}, such as text length, use of linguistic markers, or references to external links/theories---influence the response satisfaction \cite{ahmad2018survey, barua2014developers, naderi2020similarity, park2016consumers}. The second, \textbf{knowledge coverage}, evaluates the extent to which visualization-related knowledge and concepts are included. Prior studies have shown that the breadth of knowledge concepts  are important factors to ensure the qualification of knowledge exchange in specific areas of inquiry such as medical/health \cite{naderi2020similarity,park2016consumers} or software engineering \cite{ahmad2018survey,barua2014developers}. The third, \textbf{perceived quality}, assesses how participants rated the adequacy and usefulness of the responses across multiple dimensions. Recent LLM-based studies suggest that human assessments, also referred to as holistic evaluation of automatically generated texts, provide more nuanced quality of text over intrinsic properties such as coverage, depth, or actionability than evaluations based solely on automatic metrics \cite{liang2022holistic, awasthi2023humanely, sreedhar2024canttalkaboutthis,van2021human, van2018measuring}.

To operationalize these features, we employ the following methods by extracting features from texts and eliciting human perception and satisfaction. First, we computed rhetorical style features by automatically counting quantifiable attributes such as response length and by matching keywords to identify linguistic structures and explanatory styles, based on a predefined list of keywords (e.g., on the other hand, contrary to).

Second, to assess the knowledge coverage, we examined how visualization concepts in a well-known visualization model \cite{munzner2009nested} appear in the key ideas and advice of the responses. We expanded four components of visualization design into keywords to conduct keyword matching, by making it a more comprehensive list of keywords using LLMs---specifically, GPT-4o-mini, released in May 2024---through a two-step procedure below: 

\vspace{-0.75em}
\begin{itemize}
    \item We expanded the original visualization model into a three-level hierarchical taxonomy with more comprehensive collection of 587 keywords through a targeted prompting as presented in Appendix \ref{sec:appendix1}. Two human reviewers validated the expanded taxonomy through an iterative validation process.
    \vspace{-0.75em}
    \item Using the refined model, we first extracted advice segments---portions of responses offering direct recommendations, explanations, or suggestions---to filter out irrelevant content and then identified visualization concepts within these segments. We leveraged LLM capabilities with prompts detailed in Appendix \ref{sec:appendix2} to support both steps. To ensure reliability, we validated 10\% of randomly selected responses, confirming that the extracted advice accurately reflected the respondents' feedback and intent.

\end{itemize}
\vspace{-0.75em}

Lastly, to evaluate the perceived quality of responses from \h and \cg, we conducted a survey collecting human evaluations across seven metrics: coverage, breadth, topicality, depth, clarity, actionability, and overall quality. The evaluation metrics followed prior work~\cite{kim2023good}. In the survey, participants assessed a data visualization question alongside two responses---one from \cg and one from a \h---using a comparative five-point Likert scale for each metric. Participants also provided open-ended explanations for their ratings of overall quality.

We recruited 210 participants through Prolific, assigning them to one of two comparative conditions: \cgthree vs. \h or \cgfour vs. \h. Participants evaluated 35 questions randomly sampled from a pool of 119, with sampling determined through a power analysis using McNemar's test ($\alpha$ = 0.05, medium effect size, power = 0.80). For each condition, we collected three human responses per question to capture evaluation variability. Participants were compensated \$1.90 (\$13.67/hr) for completing the 8-minute survey.

\subsection{Statistical and regression analysis} 
To address RQ1 (differences in response characteristics, including rhetorical styles and knowledge coverage) and RQ2 (differences in perceived response quality), we performed statistical comparisons between \cg and \h responses. To address RQ3 (factors associated with higher quality responses), we conducted a regression analysis predicting overall quality ratings. We employed Elastic Net regression to mitigate overfitting due to feature correlations and to highlight the predictors most strongly associated with perceived quality.
\section{Findings}
\label{sec:findings}

Our analysis results reveal several key differences in multifaceted capabilities when comparing \cg and \h performance.

\paragraph{\cgfour~more closely matches \h~rhetorical styles and knowledge coverage.}
\label{sec:finding-1}
In response to RQ1, our comparative analysis shows that \cgfour aligns more closely with \h in both knowledge coverage and rhetorical style than \cgthree, while the two \cgs have a common ground that makes them distinct from \h, as detailed below (Figure \ref{fig:features}).

\textbf{Rhetorical style.} As shown in Figure \ref{fig:features}-i, the overall responses differed in their length, where \cgthree responses were twice as long as those of \cgfour and \h. On the other hand, in the analysis of normalized features (i.e., features per unit word count), \cgfour and \h exhibited higher lexical diversity (i.e., \# unique keywords) and more complex sentence structure than \cgthree. Their responses showed much dense representation of ideas in a short span of texts than \cgthree, especially for the case of \h responses with the advice-to-word ratio significantly higher than the others.

\begin{figure*}
\centering
\includegraphics[width=.9\textwidth]{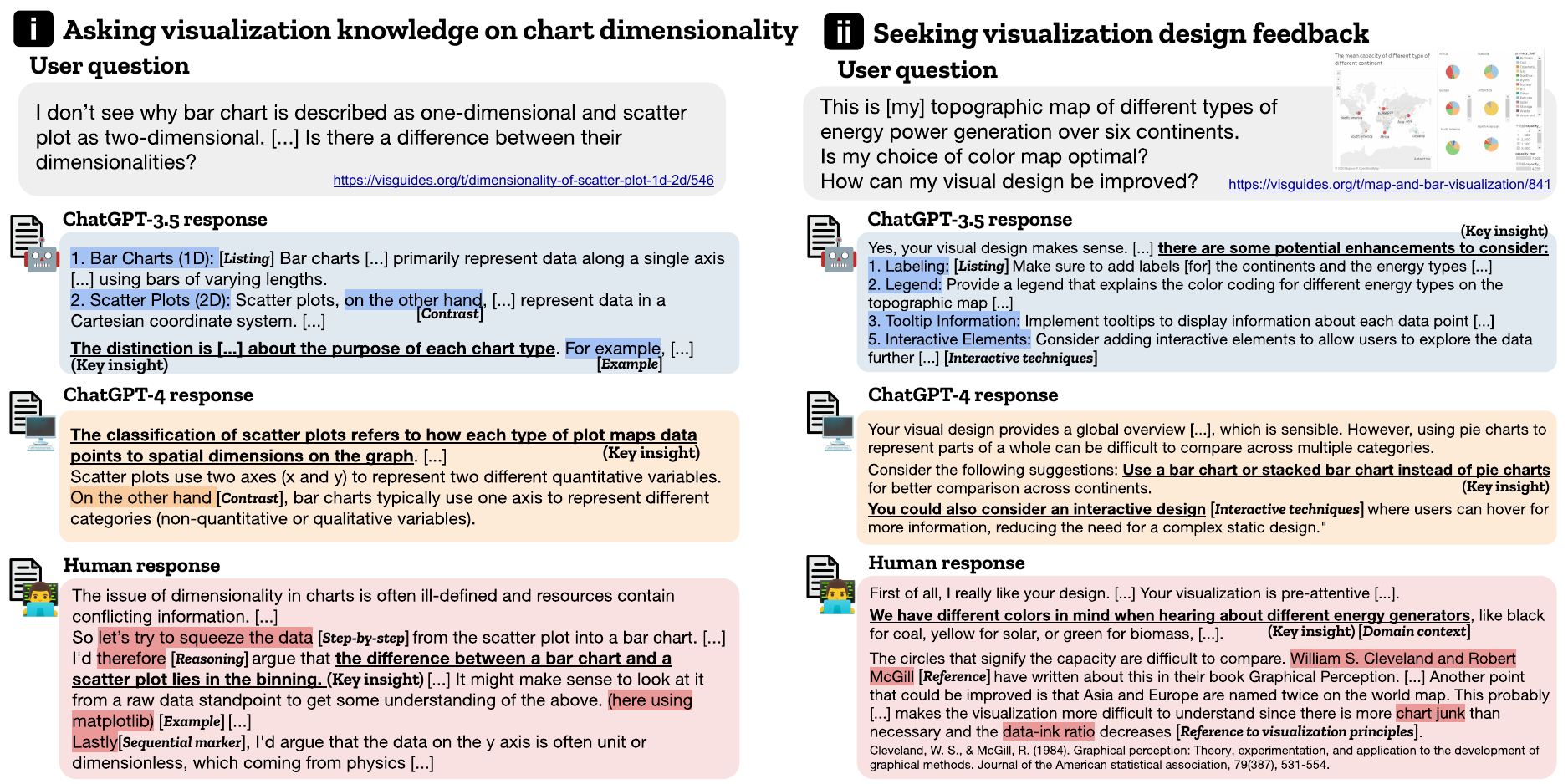}
    \vspace{-1em}
    \caption{\label{fig:case-study}
    The case study highlights two representative cases of visualization-related user question and responses from three different sources, including i) asking visualization knowledge and ii) seeking visualization design feedback. In the examples, they are distinct from each other in terms of their key insights and feedback (as underlined) and rhetorical and knowledge characteristics (as highlighted with color).}
\end{figure*}

When it comes to explanatory strategies, on the other hand, two \cgs were distinct from the \h counterpart in ways they described their feedback. For instance, \h responses tended to use more references and information sources compared to two versions of \cgs with almost no references. The use of sequential markers (e.g., first, lastly) was observed exclusively in \h responses. In contrast, \cgs frequently employed contrastive language to highlight differences as shown in Figure \ref{fig:case-study}-i. 

While mostly in common, two \cgs also exhibited different styles, where \cgthree more commonly relied on examples or bullet-point-like listings (see the example in Figure \ref{fig:case-study}-i).

\textbf{Knowledge coverage.} In the analysis of the knowledge representation (Figure \ref{fig:features}-ii), we found that two \cgs exhibit a similar span of visualization knowledge to that of \h, given the distribution of the visualization concept ratio that did not differ significantly between them. From the Chi-square analysis, \h placed significantly more emphasis on domain problems (i.e., understanding data characteristics), whereas \cgs more frequently addressed generic interaction techniques such as detail-on-demand in tooltips or drill-down views than \h does. However, when measuring the distance from the knowledge distribution of \cgthree and \cgfour to that of \h, \cgfour’s knowledge span was closer to human than that of \cgthree, indicating that \cgfour better approximates human-like knowledge coverage.

\paragraph{\cg responses are generally more preferred over human responses.}
\label{sec:finding-2}

In response to RQ2 (Figure \ref{fig:features}-iii), we found that both \cgs obtained higher scores across all intrinsic qualities than \h (all scores $>$ 3). \cg-generated responses were perceived better than human responses for all metrics, especially by larger margins in coverage, breadth, and overall quality. Participants in their open responses mentioned that, despite the strengths of human responses such as including ``\textit{someone's experience rather than a textbook explanation}’’ with ``\textit{targeted and specific tips}’’ in ``\textit{natural}’’ tones, they lack ``\textit{depth and enough information}’’ to make a clear decision, as well as ``\textit{non-user-friendly structure without numbered points}’’ and often ``\textit{go off the track}'' with information irrelevant to the questions.

Despite \cgs being preferred over \h, two versions of \cgs demonstrated distinct strengths in perceived quality when compared against each other. Specifically, \cgthree was highly rated in coverage, breadth, and depth better than \cgfour, which is in relation to its extensive description of their responses. Participants mentioned in their open responses that \cgthree ``\textit{covers different possibilities and gives the benefits of each of them}’’ (coverage) as well as ``\textit{offers exceptional coverage}'' (breadth) and ``\textit{provides [structured] guides of explanation in detail}’’ (depth). On the other hand, \cgfour's overall quality was higher than \cgthree's, particularly in terms of topicality, clarity, and actionability. as \cgfour was found to ``\textit{make it much easier to stay focused and on topic with better structure of the answer}''  (clarity and topicality) and ``\textit{presents alternative visual cues}’’ (actionability).

\begin{figure}
    \includegraphics[width=\columnwidth]{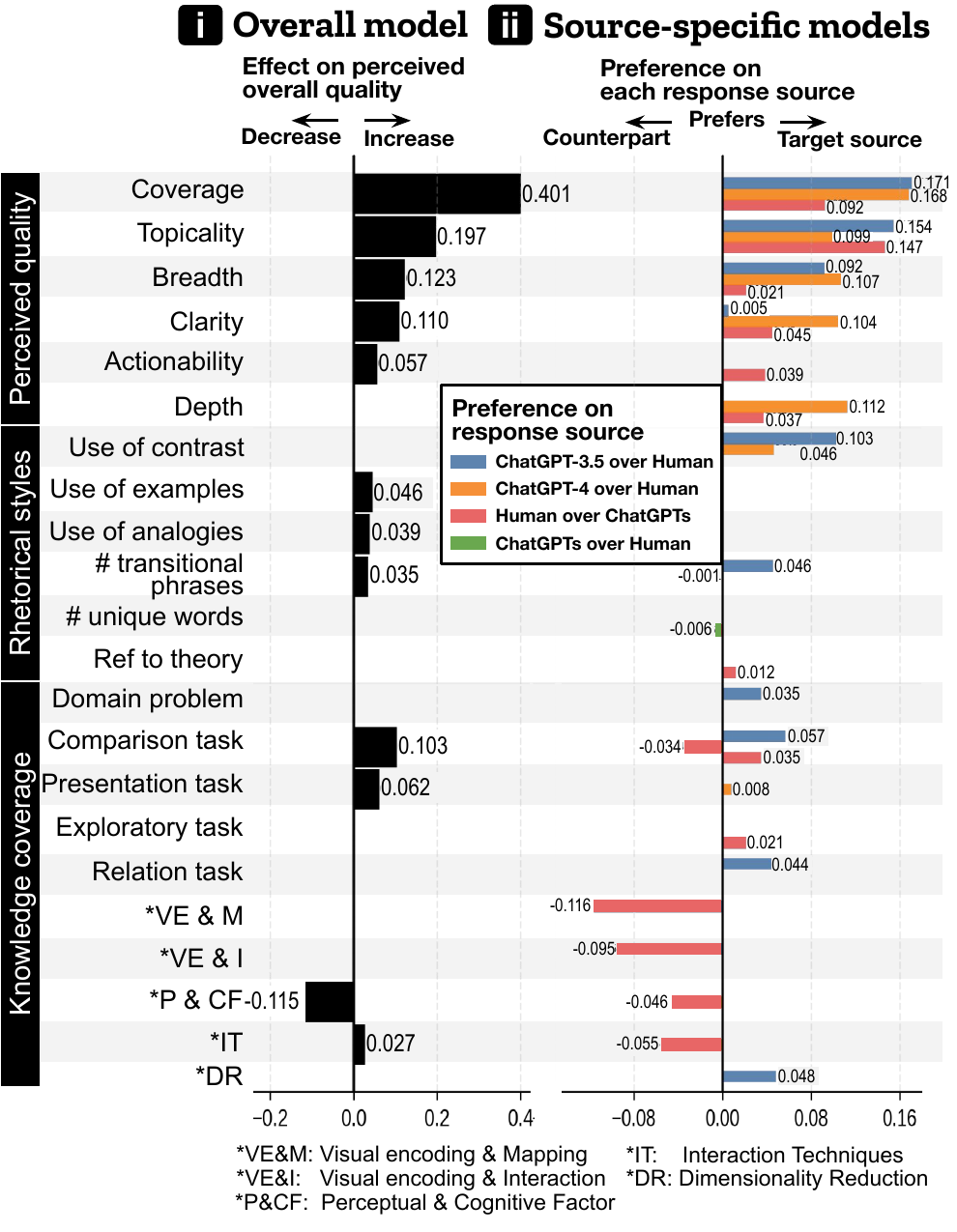}
    \vspace{-2.65em}
    \caption{\label{fig:coefs}
    The regression analysis identifies major factors associated with i) users' general preferences and ii) their preferences in favor of certain response source.}
    \vspace{-2em}
\end{figure}

\paragraph{Multiple characteristics of responses collectively shape the overall preferences.}

Two regression models---one across all response sources and another within each response source---reveal what factors collectively shape (1) users' general preferences over certain responses and (2) their preferences in favor of certain response sources over others.

\textbf{Factors associated with general preferences.}
From the overall model (Figure \ref{fig:coefs}-i), 12 factors were identified to highly influence user perception of the overall quality of responses across all response sources. Specifically, coverage and topicality were identified as the most dominantly impactful factors, indicating that people anticipate a feedback response to the point and cover all aspects of the inquiry. In addition, participants were in favor of certain types of knowledge, primarily on the higher-level feedback such as comparison and presentation tasks rather than other visualization concepts. Regarding explanatory strategies, responses were generally perceived as more useful when including examples and analogies.

\textbf{Factors associated with preferences for specific response sources.} 
\label{sec:finding-3-2}
The results from the response-source-specific models (Figure \ref{fig:coefs}-ii) revealed how each source (\textsc{\cgthree/4} or \h) was comparatively more preferable than its counterpart (\h or \cg).  In terms of perceived quality, two \cg models' coverage and breadth showed stronger associations with overall quality than those of \h. Meanwhile, \cgfour and \h's clarity and depth were more closely linked to higher quality ratings than \cgthree, suggesting that users value different strengths from each response source.

Furthermore, each response source demonstrated a distinct association between its knowledge, rhetorical style, and the perceived overall quality. Consistent with two \cgs' characteristics described in Section \ref{sec:findings}, their frequent use of contrast in explanatory styles, along with a greater focus on technique-oriented visualization concepts—relation task and dimensionality reduction for \cgthree and presentation task for \cgfour—were associated with users' higher perceived quality of responses. Human responses, on the other hand, were rated more favorably due to more references to theoretical concepts and certain visualization knowledge, especially interaction techniques, perceptual \& cognitive factors, and visual encoding. On the other hand, a number of features regarding information quantity, other stylistic and knowledge features were found to be insignificant in influencing the overall quality.

\section{Case Study}

In this section, we demonstrate how the differences among the three sources---regarding the multidimensional characteristics of their responses discussed in Section \ref{sec:findings}---are reflected in actual texts, using two specific visualization-related Q\&A examples. In these examples, we also highlight how they convey distinct key insights, which go beyond differences in rhetorical style and knowledge.

The first case (Figure \ref{fig:case-study}-i) highlights how three sources exhibit differences in their perspectives and styles in answering a knowledge-oriented visualization query—why bar charts and scatter plots, even though both use x and y axes, are considered one-dimensional and two-dimensional, respectively. As featured earlier (Figure \ref{fig:features}-i), two \cg models commonly \textbf{use contrast} to highlight differences between the two chart types. However, in their key insights, \cgthree emphasizes different purposes of each chart type (e.g., comparing values along a single dimension vs. examining relationships between two variables), while \cgfour highlights how quantitative variables are mapped to spatial dimensions, linking this to the notion of dimensionality. the human response, while aligned with \cgfour’s interpretation in highlighting data structure and encoding, offers a \textbf{reasoning process} of how two plots fundamentally differ from each other \textbf{in a step-by-step manner}. Starting by acknowledging the ambiguity in defining dimensionality in charts, it presents a key perspective that the distinction lies in binning, in a step-by-step walk-through with Python code and references to a data analysis tool.

In the second example (Figure \ref{fig:case-study}-ii), the three response sources also took noticeably different approaches, key ideas, and focal knowledge in providing feedback regarding the use of a pie chart, color choices, and the overall layout of a composite interface. As highlighted in Figure \ref{fig:features}-i, the human response stood out for its \textbf{theoretical depth, referencing perceptual principles} such as pre-attentive processing or data-ink ratio, along with citations for graphical perception and domain-specific insights about color choice with respect to the types of energy generator. Compared to \h, two \cg responses were grounded in \textbf{practical solutions} by commonly making an emphasis on interaction techniques, as previously highlighted in \ref{fig:features}-ii, but with approaches quite distinct from each other. \cgthree \textbf{took a conservative stance}, recommending an incremental improvement---such as clearer labeling, consistent color coding, and interactive elements like tooltips. \cgfour, on the other hand, \textbf{proposed more fundamental changes}, suggesting an alternative chart type (e.g., stacked bar charts) and the inclusion of interactive features.
\section{Discussion \& Conclusion}
\label{sec:discussion}
In this work, we investigated why and how ChatGPT has better capabilities than human in giving visualization feedback. Based on findings, we draw implications for the potential of LLMs and human perception over capabilities over broader applications of AI.

First, our analysis shows that LLMs provide solid visualization feedback and make progress toward combining the strengths of both human feedback and their own capabilities. As observed with \cgfour, the model demonstrated both extensive knowledge of machine intelligence and human-like text generation. While this indicates that LLMs can serve as alternatives to human feedback, humans are still distinct from LLMs and outperform in some dimensions such as domain-specific problem knowledge and heuristics, step-by-step guidance, and references to theoretical frameworks. Thus, exploring ways in
which humans and AI can complement each other may yield better knowledge outcomes or offer a pathway to further advance LLMs
by learning from how humans articulate their intent and feedback.

Furthermore, our regression analysis revealed that human evaluation is highly discriminative—users do not simply reward verbosity or surface-level traits, but instead respond to specific types of knowledge and explanatory strategies. This finding sheds light on the properties that enable chat-based services to deliver generally satisfying responses, particularly by designing prompt strategies that incorporate effective explanatory techniques. In addition, we observed substantial variation in individual user preferences regarding style, knowledge depth, and quality dimensions. These insights highlight the need for personalization; for example, some users may prefer concise lists of action items, whereas others may find detailed explanations and references to specific knowledge more informative. We find that enhancing user experience may be achieved by designing systems that present a range of response options explicitly or that motivate users to refine their queries or encourage follow-up prompts through some strategies such as cognitive prompts or nudging techniques. In addition, future work can enhance the generalizability of our analysis by leveraging larger datasets to control quality variations as well as advanced LLM models, and conducting task-based evaluations.

\section{Acknowledgement}
We would like to acknowledge the support of the National Science Foundation (\#2146868).


\bibliographystyle{abbrv-doi}

\bibliography{references}

\appendix

\section{Prompts for keyword expansion on visualization taxonomy}
\label{sec:appendix1}
\begin{small}
\begin{spverbatim}
This classification framework organizes visualization knowledge into three hierarchical levels: Level 1 (L1) and Level 2 (L2) denote conceptual categories, while Level 3 (L3) contains associated keywords. Expand the L3 keyword set to improve coverage and enable the detection of visualization-related concepts within text.
\end{spverbatim}
\end{small}

\section{Prompts for advice identification and visualization concept detection from text}
\label{sec:appendix2}
\begin{small}
\begin{spverbatim}
The given text is a response to a visualization-related question.
    
Conduct the analysis in the following steps:
  1) Identify unique pieces of advice (title and description),
  2) For each advice, detect multiple visualization-related keywords as concepts that appear in the Categories (L1 and L2) attached below, each with L1 and L2 category and phrases/sentences as evidence on which keywords appear.

As an output, present a list of pairs of visualization advice:(advice_title, advice_description, L1, L2, L3 (keyword), evidence) with the references of Taxonomy in a json format.

    Text: {},
    Categories (L1 and L2): {},
    Taxonomy: {}
\end{spverbatim}
\end{small}

\end{document}